\documentclass[fleqn,usenatbib]{mnras}

\usepackage{newtxtext,newtxmath}
\usepackage[T1]{fontenc}

\DeclareRobustCommand{\VAN}[3]{#2}
\let\VANthebibliography\thebibliography
\def\thebibliography{\DeclareRobustCommand{\VAN}[3]{##3}\VANthebibliography}

\usepackage{graphicx}
\usepackage{amsmath}
\usepackage[english]{babel}
\usepackage{tabularx}
\usepackage{caption}
\usepackage{subcaption}
\usepackage{multicol}
\usepackage{placeins}

\title[Convective vortices in collapsing stars]{Convective vortices in collapsing stars}

\author[Telman et al.]{
Yerassyl Telman,$^{1} \thanks{Email: yerassyl.telman@alumni.nu.edu.kz}$
Ernazar Abdikamalov,$^{1,2} \thanks{Email: ernazar.abdikamalov@nu.edu.kz}$
Thierry Foglizzo$^{3}$ \thanks{E-mail: foglizzo@cea.fr}\\
$^{1}$Department of Physics, Nazarbayev University, 53 Kabanbay Batyr ave, 010000 Astana, Kazakhstan \\ 
$^{2}$Energetic Cosmos Laboratory, Nazarbayev University, 53 Kabanbay Batyr ave, 010000 Astana, Kazakhstan\\
$^{3}$AIM, CEA, CNRS, Universit\'e Paris-Saclay, Universit\'e Paris Diderot, Sorbonne Paris Cit\'e, F-91191 Gif-sur-Yvette, France}

\date{Accepted XXX. Received YYY; in original form ZZZ}

\pubyear{\the\year{}}

\begin{document}
\label{firstpage}
\pagerange{\pageref{firstpage}--\pageref{lastpage}}
\maketitle

\begin{abstract}
Recent studies show that non-radial structures arising from massive star shell convection play an important role in shaping core-collapse supernova explosions. During the collapse phase, convective vortices generate acoustic waves that interact with the supernova shock. This amplifies turbulence in the post-shock region, contributing to explosion. We study how various physical parameters influence the evolution of these convective vortices during stellar collapse using simplified simulations. We model the collapsing star with a transonic Bondi flow and represent convection as solenoidal velocity perturbations. Our results are consistent with previous studies, demonstrating that the peak perturbation amplitude scales linearly with the pre-collapse convective Mach number and inversely with the angular wavenumber of convection. While the shell radius and width primarily determine the timescale of accretion, they have little impact on the peak perturbation amplitudes. Finally, we show that when the convective Mach number is below approximately 0.2, the dynamics remain within the linear regime.
\end{abstract}



\section{Introduction}
\label{sec:intro}

It is well established that the radial structure of massive stars plays a crucial role in determining the outcomes of the core-collapse supernova (CCSN) explosions they may undergo \cite[e.g.,][]{oconnor11black, ugliano12progenitor, suwa16Acriterion, ertl16two, bruenn16development, burrows19three}. Recent studies show that non-radial structures caused by shell convection can significantly influence these supernova explosions as well \cite{couch13revival, takahashi16links, radice17electron}. When these perturbations accrete through the supernova shock, they amplify the turbulence in the post-shock region, which aids the exlosion by pushing the shock forward \cite[e.g.,][]{muller15non}.

In recent years, 3D simulations of the final stages of core-collapse supernova (CCSN) progenitors have become available \cite[e.g.,][]{couch15three, yadav20large, yoshida21three, fields21three, varma233D, georgy243D}. These simulations confirm the presence of strong convection within the innermost nuclear burning shells, especially those involving silicon and oxygen burning. Many progenitors, especially those with initial masses between $\sim 15$ and $\sim 25$ solar masses, exhibit large-scale convective modes in their oxygen-burning shells, which are expected to have a particularly significant impact on the CCSN explosion dynamics \cite{collions18properties}.

The evolution of convective vortices as they fall inward during  stellar collapse is complex. As these vortices descend \cite{takahashi14linear}, they distort iso-density surfaces, resulting in density perturbations \cite{foglizzo01entropy}. These perturbations generate pressure fluctuations that propagate as acoustic waves \cite{muller16last}. At the same time, the accelerated flow causes the vortices to stretch in the radial direction. Given that circulation, defined as the line integral of velocity around the vortex, is conserved \cite{landau87fluid}, this stretching results in a substantial decrease in vortex velocity as they reach smaller radii \cite{abdikamalov20acoustic}. Consequently, primarily acoustic waves are able to penetrate into the inner regions of the flow.

While these perturbations travels toward the center, other key processes occur in the inner regions. The collapse of the iron core launches a shock wave. For a supernova explosion to occur and leave behind a stable neutron star, this shock must expel the stellar envelope \cite[e.g.,][]{janka01conditions}. However, as the shock propagates outward, the dissociation of heavy nuclei and neutrino cooling drain its energy, causing it to stall at $\sim 150 \, \mathrm{km}$ within milliseconds of its formation \cite[e.g.,][]{liebendoerfer01probing}. Neutrinos emitted by the newly formed protoneutron star heat the post-shock flow, driving neutrino-driven turbulent convection \cite[e.g.,][]{herant94inside, burrows95on, janka95first}. The turbulent pressure helps push the shock forward \cite{murphy13dominance, radice16neutrino, melson20resolution}. Additionally, if present, the standing accretion shock instability can also contribute to the explosion \cite{blondin03stability, foglizzo07sasi, iwakami08three, fernandez15three}. Magnetic fields may also play a role in the explosion dynamics \cite{endeve12turbulent, matsumoto22magnetic, varma233D}. See \cite{burrows13colloquium, janka16physics, mezzacappa20physical} for recent reviews. 

The acoustic waves originating from the nuclear burning convective shell encounter the stalled supernova shock within a few hundred milliseconds after core collapse \cite[e.g,][]{muller17supernova}. This encounter disturbs the shock \cite{abdikamalov16shock}, generating entropy perturbations in the post-shock region \cite{huete18impact, abdikamalov18turbulence}. The buoyancy of these perturbations drives non-radial flows, enhancing the convection behind the supernova shock \cite{kazeroni20impact, muller16last}. This pushes the shock forward and aids the explosion \cite[e.g.,][]{radice18turbulence}.

In this work, we investigate the evolution of convective vortices during stellar collapse through a parameter study using simplified simulations. We model the stellar collapse using a transonic Bondi flow \cite{bondi52spherically} with an imposed solenoidal velocity field that mimics convective motion. We conduct a series of simulations, varying the angular wavenumber $\ell$, the shell radius and width, as well as the convective Mach number $\delta \cal{M}$. Our setup does not include the supernova shock or the protoneutron star. The main advantage of this simplified approach is that we can separately study the impact of different physical parameters with moderate computational cost, which is hard to achieve with detailed multi-physics and multi-scale 3D radiation-hydrodynamics simulations \cite[e.g.,][for a recent review]{muller20hydrodynamics}. Our work thus complements detailed numerical simulations \cite[e.g., ][]{muller15non}, while bridging the gap with and verifying the validity of our previous linear perturbative studies \cite[e.g.,][]{abdikamalov20acoustic, abdikamalov21impact}.

This paper is organized as follows: Section \ref{sec:methods} outlines our methods, Section \ref{sec:results} presents our findings. Finally in Section \ref{sec:conclusion}, we present our conclusions.

\section{Methods}
\label{sec:methods}

\begin{figure}
    \centering
    \includegraphics[width=0.5\textwidth]{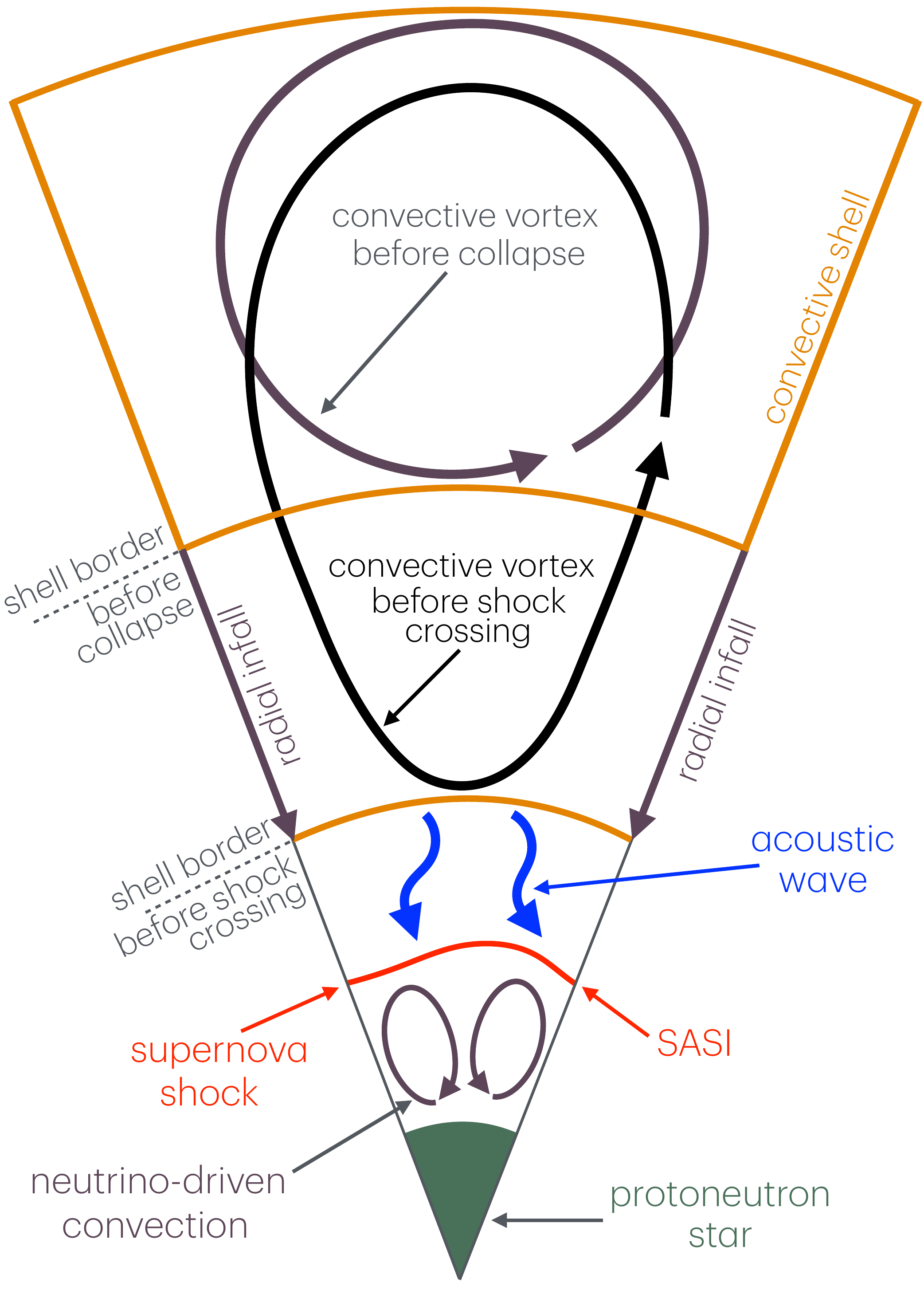}
    \caption{A schematic illustration of convective vortex dynamics within a collapsing star. Convective vortices distort the isodensity surfaces, generating pressure disturbances that propagate as acoustic waves. These waves interact with the supernova shock, leading to the formation of entropy perturbations in the post-shock region. The buoyancy of with these perturbations induces non-radial motion, which, along with neutrino-driven convection and the standing accretion shock instability, contributes to the forward progression of the shock.}
    \label{fig:eddy_dynamics}
\end{figure}

To model the evolution of convective vortices, we employ the hydrodynamics finite-volume code {\tt IDEFIX} \cite{lesur23idefix}. We utilize the HLL Riemann solver and the third-order Lim03 reconstruction \cite{vcada2009compact}. For verification, we repeat a subset of simulations using second-order piecewise linear reconstruction (PLM) with the Van-Leer slope limiter, and find nearly identical results. Given that pre-collapse convection is subsonic \cite[e.g.,][]{collions18properties}, the convective turnover timescale is longer than the collapse timescale, which proceeds at supersonic velocities. Consequently, turbulent energy transfer to smaller scales is unlikely to affect the dynamics at the largest scale during collapse. The same holds true for the impact of nuclear burning and neutrino cooling. Therefore, we assume an adiabatic evolution. We model stellar matter as an ideal gas with an adiabatic index $\gamma=4/3$.

For the vast majority of stars, the rotation is expected to be too slow to play any significant role in collapse dynamics \cite[e.g.,][]{heger05presupernova, popov12, noutsos13, cantiello14angular, deheuvels14seismic}. We thus consider non-rotating stars.  The collapsing star is modeled via the transonic Bondi solution \cite{bondi52spherically}. This flow exhibits a sonic radius $r_\mathrm{s}$, below (above) which the flow is supersonic (subsonic), which mimics the collapse of real stars before explosion sets in. 

The outer boundary of the computational domain is located at twice the outer shell boundary radius of the model with the largest shell (details on model parameters provided below), we impose a flow corresponding to the Bondi solution. The inner boundary is located at $0.0125 R_\mathrm{shell}$, where $ R_\mathrm{shell}$ is the initial shell radius of our standard model. At this boundary, we apply the outflow boundary condition. Since the convective shell has a radius of a few thousand kilometers before core collapse \cite{muller16last}, the inner boundary ($0.0125 R_\mathrm{shell}$) is below the $\sim 150$ km region where the stalled shock is expected to reside. Consequently, the computational domain is sufficiently wide to capture the evolution of convective vortices from their location before core collapse to the point where they encounter the supernova shock. We conduct two-dimensional simulations under the assumption of axial symmetry. We employ 250 logarithmic radial cells and 80 equidistant angular cells. Our extensive resolution test suggests that this resolution is sufficient for accurately capturing all the physical processes studied below. 

To model convection within this flow, we introduce velocity perturbations between the inner and outer radii of the convective shell, $r_{\text{min}}$ and $r_{\text{max}}$. Following \cite{muller15non}, these perturbations are defined as
\begin{equation}
\delta \mathbf{v} = \frac{C}{\rho} \nabla \times \psi,
\end{equation}
where $C$ is a normalization factor and the function $\psi$ is given by
\begin{equation}
\psi = \mathbf{e_\varphi} \frac{\sqrt{\sin \theta}}{r} \sin \left(\pi \frac{r - r_{\text{min}}}{r_{\text{max}} - r_{\text{min}}} \right) Y_{\ell,1}(\theta, 0).
\end{equation}
This formulation ensures that the velocity field is solenoidal, a characteristic of the shell convection in massive stars \cite{muller15non}. The inclusion of the factor $\sqrt{\sin \theta}$ and the use of spherical harmonics $Y_{\ell,1}$ (instead of $Y_{\ell,0}$) are done to avoid singularities at $\theta = 0$ and $\theta = \pi$. This approach also guarantees vanishing non-radial velocities along the coordinate axis and ensures that radial velocities are zero at both the inner and outer boundaries of the convective shell. While this method does not capture the full spectrum of turbulent motion, an area where the approach by \cite{chatzopoulos14characterizing} may be more suitable and which lies beyond the scope of this work, it offers the advantage of easily generating a velocity field with a given spatial scale, making it convenient for parameter studies \cite{muller15non}.

We explore a range of models by varying the angular wavenumber $\ell$, average shell radius $R_\mathrm{shell}$, shell width $\Delta R_\mathrm{shell}$, and convective Mach number $\delta {\cal M}$. Our standard model is characterized by $\delta {\cal M} = 0.1$, $\ell = 2$, $R_\mathrm{shell} = 2r_\mathrm{s}$, and $\Delta R_\mathrm{shell} = \pi R_\mathrm{shell} / (2\ell)$. To assess the impact of these parameters, we generate sequences of models where only one parameter is varied while the rest are held constant, matching the standard model.

In the first sequence, we vary the angular wavenumber, considering $\ell = 1$, $2$, $4$, and $8$. In the second sequence, we explore different shell radii with $R_\mathrm{shell} = 0.5 r_\mathrm{s}$, $r_\mathrm{s}$, and $2 r_\mathrm{s}$. For the model with $R_\mathrm{shell} = 0.5 r_\mathrm{s}$, the convective shell is initially contained within the supersonic region, whereas for $R_\mathrm{shell} = 2 r_\mathrm{s}$, it is within the subsonic region. In the model with $R_\mathrm{shell} = r_\mathrm{s}$, the inner half of the convective shell is initially in the supersonic region, while the outer half is in the subsonic region. In the third sequence, we adjust the shell width, ranging from $\Delta R_\mathrm{shell} = \pi R_\mathrm{shell} / \ell$ to $\Delta R_\mathrm{shell} = \pi R_\mathrm{shell} / (4\ell)$. Finally, in the fourth sequence, we vary the convective Mach number $\delta {\cal M}$ from $0.05$ to $0.4$. Hereafter, time is measured in units of $t_\mathrm{adv}$, the time for the inner shell boundary to cross the computational domain. All models evolve up to $t/t_\mathrm{adv} \approx 21.78$ for the standard model, which is roughly $17.25$ times the advection timescale $r_\mathrm{s} / c_\mathrm{s}$ through the sonic radius, where $c_\mathrm{s}$ is the speed of sound at $r_\mathrm{s}$. 

\section{Results}
\label{sec:results}

\begin{figure*}
    \centering
    \includegraphics[width=\textwidth]{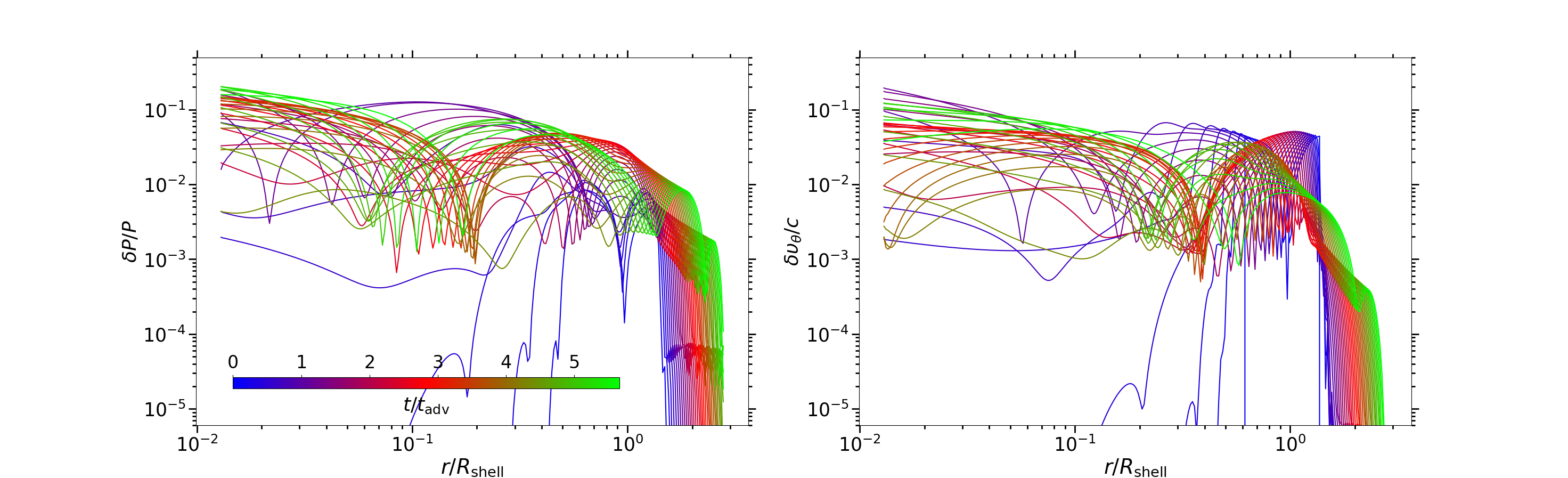}
    \caption{The angle-averaged radial profiles of $\delta P/P$ (left panel) and $\delta \upsilon{\!}_{\theta}/c$ (right panel) at different times for the standard model ($\ell = 2$, $R_\mathrm{shell}=2 r_\mathrm{s}$, $\Delta R_\mathrm{shell} = {\pi R_\mathrm{shell}}/{(2 \ell)}$, $\delta \mathcal{M}=0.1$). The color of each line corresponds to the value of $t/t_\mathrm{adv}$, where $t_\mathrm{adv}$ is the accretion time of the inner shell boundary through the inner edge of the computational domain at $0.0125 R_\mathrm{shell}$.  The last time snapshot marks the moment the outer boundary of the convective shell accretes through the inner edge. The amplitudes of both pressure and velocity perturbations saturate at approximately $0.1$ at small radii, which roughly corresponds to the value of convective Mach number $\delta {\cal M}$.}
    \label{fig:rad_profile}
\end{figure*}

\begin{figure*}
    \centering
    \includegraphics[width=\textwidth]{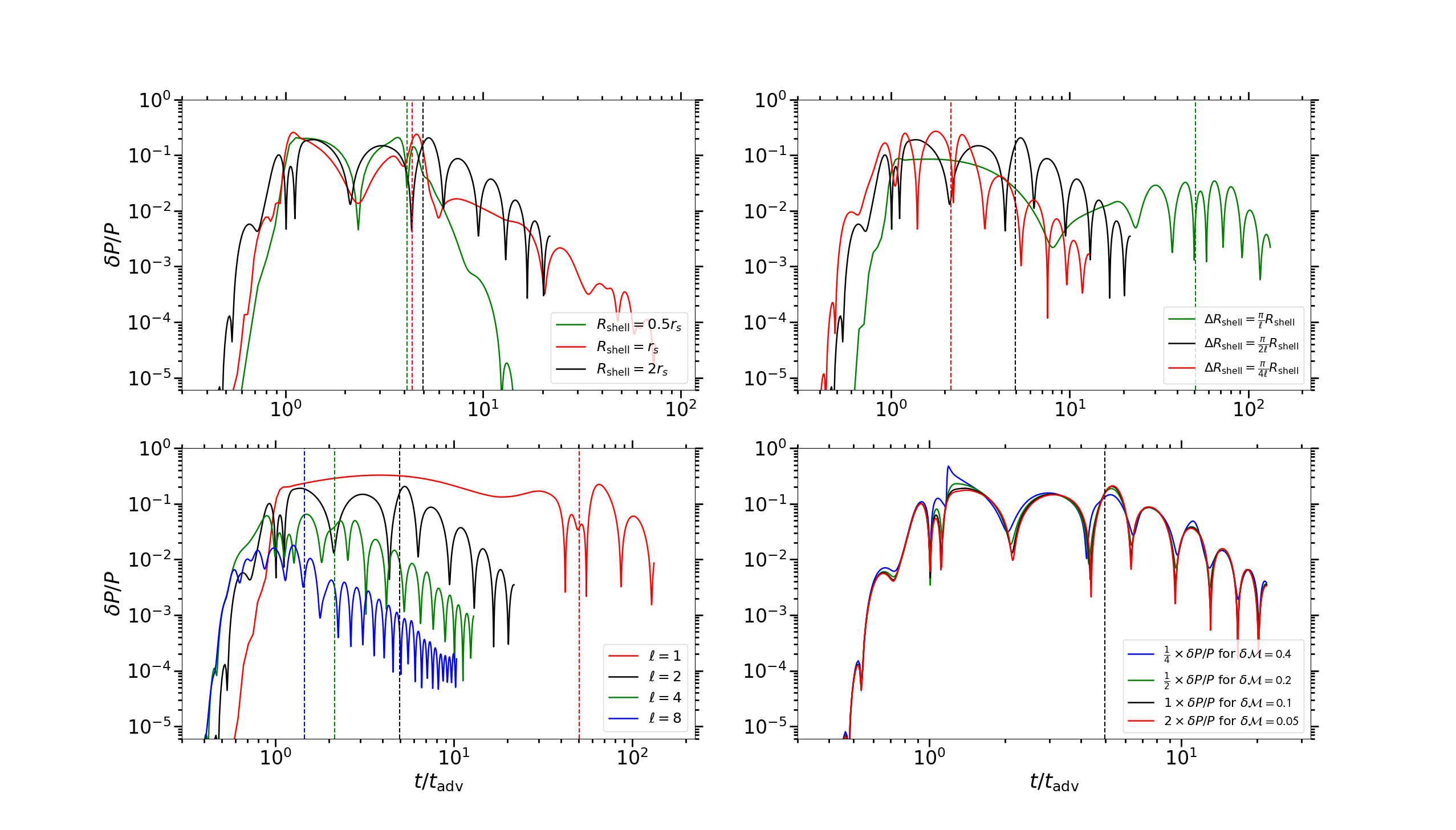}
    \caption{The angle-averaged values of ${\delta P}/{P}$ at the inner boundary of the computational domain ($r = 0.0125 R_\mathrm{shell}$) are shown as a function of time for  shell radius $R_\mathrm{shell}$ (top left), shell width $\Delta R_\mathrm{shell}$ (top right), $\ell$ (bottom left), and convective Mach number $\delta {\cal M}$ (bottom right). The dashed vertical lines indicate the moment when the outer boundary of the convective shell reaches the inner edge of computational domain. The time $t_\mathrm{adv}$ marks the accretion of the inner shell boundary through this radius.}
    \label{fig:dP_seq}
\end{figure*}

The qualitative evolution of convective vortices during stellar collapse unfolds as follows. Consider a convective vortex descending within the collapsing star. As a fluid element is displaced by a radial distance $\delta r$ due to convective motion, it induces density perturbations by distorting the iso-density surfaces of the star:
\begin{equation}
    \frac{\delta \rho}{\rho} \sim \frac{\partial  \ln \rho}{ \partial \ln r} \frac{\delta r}{r}.
\end{equation}
These density variations, in turn, generate pressure perturbations:
\begin{equation}
    \frac{\delta P }{P} \sim \frac{\delta \rho}{\rho}.
\end{equation}
These perturbations propagate as acoustic waves. This process in depicted schematically in Fig.~\ref{fig:eddy_dynamics}. The displacement $\delta r$ is related to the radial velocity perturbation $\delta \upsilon_r$ by $\delta r \sim {\delta \upsilon_r}/{\omega}$, where $\omega$ is the angular velocity. The radial velocity $\delta \upsilon_r$ is connected to vorticity perturbations $\delta \omega$ via the relation $\delta \omega \sim {i m \delta \upsilon_r}/{r}$, where $m$ is the angular wavenumber in the $\phi$ direction. By combining these equations, we find that
\begin{equation}
\label{eq:dp_vs_m}
    \frac{\delta P}{P} \propto \frac{1}{m},
\end{equation}
indicating that the amplitude of the acoustic waves is inversely proportional to the angular wavenumber of the convection. This scaling is consistent with the results of our linear perturbative study \cite{abdikamalov20acoustic}. It is also observed in our simulations, as we show below.

We begin by analyzing the evolution of radial profiles of perturbation amplitudes from the onset of collapse until the shell outer boundary accretes through the inner edge of our computational domain. This evolution is illustrated in Fig.~\ref{fig:rad_profile}, where the left panel shows the angle-averaged radial profile of pressure perturbations, $\delta P/P$, while the right panel shows the transverse velocity perturbations, ${\delta \upsilon{\!}_{\theta}}/{c}$, where $c$ is the speed of sound, for the standard model at various time steps. As the perturbations move inward, their amplitudes stabilize around $\sim 0.1$. The amplitude exhibits minimal variation with radius, except for the sharp dips discussed later. This suggests that when these perturbations reach the stalled supernova shock, which is expected to be located at relatively small radii (compared to the initial shell radius), the acoustic waves should maintain an amplitude of $\delta P/P \sim 0.1$, corresponding to the value of convective Mach number $\delta {\cal M}$. This is consistent with qualitative estimates \cite{muller16last} and numerical simulations \cite{muller17supernova}.

Notably, the transverse velocity, $\delta \upsilon{\!}_{\theta}$, does not follow the $\propto 1/r$ scaling \cite{kovalenko98instability, lai00growth}, which is typically observed for rotational velocity in a collapsing star due to the conservation of angular momentum. The circulation, defined as the integral of velocity around a closed curve
\begin{equation}
\Gamma=\oint {\bf \upsilon} d {\bf s},
\end{equation}
is conserved in isentropic flows \cite{landau87fluid}. As the flow accelerates, vortices stretch radially, requiring a decrease in velocity to conserve $\Gamma$, limiting ${\delta \upsilon{\!}_{\theta}}/{c}$ to $\sim 0.1$. This finding, initially observed in our linear perturbative study \cite{abdikamalov20acoustic}, is now confirmed by our current simulations. The sharp dips visible in Fig.~\ref{fig:rad_profile} arise from the interaction between ingoing and outgoing acoustic waves. The latter is generated by the refraction of the ingoing waves \cite{foglizzo02nonradial}. 

Now we examine how the dynamics depends on the model parameters. Fig.~\ref{fig:dP_seq} shows the angle-averaged value of $\delta P/P$ at the inner edge of the the computational domain, where the stalled CCSN shock is expected to encounter these perturbations\footnote{Since both $\delta P/P$ and ${\delta \upsilon{\!}_{\theta}}/{c}$ are not sensitive to $r$ in the inner regions, the precise definition of the inner boundary is not necessary.}, as a function of time $t/t_\mathrm{adv}$. In all cases, the peak value of $\delta P/P$ is reached when the inner shell boundary accretes through the inner edge of the computational domain, i.e. when $t/t_\mathrm{adv} \approx 1$. The amplitudes remain close to the peak value until the outer shell boundary accretes through that edge, which is marked by dashed vertical lines. 

The top right and left panels of Fig.~\ref{fig:dP_seq} show $\delta P/P$ for models with varying shell radii $R_\mathrm{shell}$ and shell widths $\Delta R_\mathrm{shell}$, respectively. The peak value of $\delta P/P$ is not sensitive to either $R_\mathrm{shell}$ or $\Delta R_\mathrm{shell}$. In models with larger  $R_\mathrm{shell}$ or $\Delta R_\mathrm{shell}$, the outer shell radius is located further from the center, so it takes longer for them to accrete. Hence, the perturbations persist longer. 

After the accretion of the outer shell boundary through the computational domain, the acoustic waves fall off approximately as $ \propto t^{-\alpha}$, where $\alpha$ typically ranges from $\sim 2.5$ to $\sim 3$, depending on the model. These sustained perturbations are caused by acoustic waves that are emitted in the positive radial direction. Since accretion is supersonic, these waves are unable to escape beyond the sonic surface. Instead, they eventually accrete over a timescale $\int {dr}/{(|\upsilon|-c)}$, where $\upsilon$ is the accretion velocity. As expected, in models with larger $R_\mathrm{shell}$ or $\Delta R_\mathrm{shell}$, which have larger outer boundaries, these sustained acoustic waves start later and thus persist for a longer time. In contrast, in the model with $R_\mathrm{shell} = 0.5 r_\mathrm{s}$, the shell initially resides entirely within the supersonic region. Hence, the decline is steepest, scaling as $ \propto t^{-8.4}$.

The bottom left panel shows $\delta P/P$ for different $\ell$. We find that the peak $\delta P/P$ scales as $\sim 0.1 / \ell$, which is consistent with the qualitative estimate (\ref{eq:dp_vs_m}) and our previous perturbative studies \cite{abdikamalov20acoustic}. Additionally, since the shell width scales as $\propto 1/\ell$, models with a larger $\ell$ have smaller $\Delta R_\mathrm{shell}$, so they accrete on shorter timescales.

We also assess whether the dynamics of convective perturbations remain within the linear regime during collapse. To do this, we perform simulations with different convective Mach numbers, $\delta {\cal M}$, ranging from $0.05$ to $0.4$. We test the linearity by checking for a proportional relationship between $\delta P/P$ and $\delta {\cal M}$. Some models exhibit this relationship. For example, in the model with $\delta {\cal M} = 0.05$, the $\delta P/P$ is about half of that in the model with $\delta {\cal M} = 0.1$. This linear relationship holds across all models with $\delta {\cal M} \lesssim 0.2$, as illustrated in the bottom right panel of Fig.~\ref{fig:dP_seq}. This suggests that the perturbation dynamics remain within the linear regime for these values of convective Mach numbers, which are typical for convection in the innermost shells of most stars \cite{collions18properties}. This supports the validity of the linear approximation used in our previous studies \cite{abdikamalov20acoustic, abdikamalov21impact}. The model with $\delta {\cal M} = 0.4$ exhibits deviations from the linear relationship on the order of a few percent, signaling the emergence of non-linear dynamics.  

Note that a direct comparison between the simulation results and our previous linear perturbative approach is not possible. This is because the latter is based on a stationary solution for the accretion of vorticity waves with an infinitely long duration \cite{abdikamalov20acoustic}. In contrast, in this work we consider non-stationary scenario due to the finite width of the convective shell, which better represents realistic stellar models. 

Finally, we evaluate the sensitivity of our results to the details of the initial accretion velocity profile. While the Bondi solution represents a stationary flow, realistic collapse originates from a quasi-static state. To simulate this more realistically, we conduct a test where the initial velocity is set to zero (except for the convective motion), while the density distribution corresponds to that of the stationary Bondi solution. Over a dynamical timescale, this setup naturally evolves toward the Bondi velocity profile. We find that the results remain both qualitatively and quantitatively consistent with those obtained using the stationary Bondi flow, with amplitudes differing by $\lesssim 1\, \%$. This suggests that our findings are robust and not significantly influenced by the initial velocity profile of the accretion flow.

\section{Conclusions}
\label{sec:conclusion}

In this work, we studied the evolution of convective vortices in collapsing stars using parameterized numerical simulations. We explored a wide range of models with varying convective shell radii, widths, angular wavenumbers ($\ell$), and convective Mach numbers ($\delta {\cal M}$). 

We found that the peak amplitude of perturbations, upon reaching the supernova shock, scales as $\propto \delta {\cal M} / \ell$, consistent with our previous perturbative studies \cite{abdikamalov20acoustic}. Interestingly, the peak amplitudes are largely insensitive to the initial radius and width of the convective shell. In agreement with the linear perturbative studies, we do not observe the $ \propto 1/r$ scaling for the transverse velocities that was found in previous studies. Non-uniform accretion velocity stretches convective vortices and the conservation of circulation limits the growth of the transverse velocities. Finally, we conducted a series of simulations with varying $\delta {\cal M}$. We find that the models with $\delta {\cal M} \lesssim 0.2$ remain in linear regime. The model with $\delta {\cal M} = 0.4$ shows deviations from linearity, indicating the onset of non-linear dynamics. 

As previously noted, our study relies on simplified numerical simulations. We modeled the collapsing star using the transonic Bondi flow and represented initial convective motions with a basic solenoidal velocity field. Our approach deliberately omits complexities such as nuclear burning, neutrino heating, and the turbulent energy cascade to smaller scales. Although these simplifications prevent us from capturing the full complexity of the phenomenon, they enable us to systematically analyze the influence of individual physical parameters at a relatively low computational cost, a task that would be challenging with comprehensive 3D neutrino-hydrodynamics simulations. Our method serves as an intermediate step between these detailed simulations and linear perturbative analyses. In future work, we plan to extend this idealized framework to explore how these perturbations interact with the supernova shock and the post-shock flow.

\section*{Acknowledgements}

This work is supported by the Nazarbayev University Faculty Development Competitive Research Grant Program (Grant No. 11022021FD2912) and the Ministry of Education and Science, Republic of Kazakhstan (Grant No. AP19677351 and AP13067834). We thank Bekdaulet Shukirgaliyev for his help with numerical simulations. We thank Anton Desyatnikov for fruitful discussion on the linear and non-linear evolution of vortices. 

\section*{Data Availability}
 
The data underlying this work is available from authors upon request.


\bibliographystyle{mnras}
\bibliography{refs}

\begin{thebibliography}{}
\makeatletter
\relax
\def\mn@urlcharsother{\let\do\@makeother \do\$\do\&\do\#\do\^\do\_\do\%\do\~}
\def\mn@doi{\begingroup\mn@urlcharsother \@ifnextchar [ {\mn@doi@} {\mn@doi@[]}}
\def\mn@doi@[#1]#2{\def\@tempa{#1}\ifx\@tempa\@empty \href {http://dx.doi.org/#2} {doi:#2}\else \href {http://dx.doi.org/#2} {#1}\fi \endgroup}
\def\mn@eprint#1#2{\mn@eprint@#1:#2::\@nil}
\def\mn@eprint@arXiv#1{\href {http://arxiv.org/abs/#1} {{\tt arXiv:#1}}}
\def\mn@eprint@dblp#1{\href {http://dblp.uni-trier.de/rec/bibtex/#1.xml} {dblp:#1}}
\def\mn@eprint@#1:#2:#3:#4\@nil{\def\@tempa {#1}\def\@tempb {#2}\def\@tempc {#3}\ifx \@tempc \@empty \let \@tempc \@tempb \let \@tempb \@tempa \fi \ifx \@tempb \@empty \def\@tempb {arXiv}\fi \@ifundefined {mn@eprint@\@tempb}{\@tempb:\@tempc}{\expandafter \expandafter \csname mn@eprint@\@tempb\endcsname \expandafter{\@tempc}}}

\bibitem[\protect\citeauthoryear{Abdikamalov \& Foglizzo}{Abdikamalov \& Foglizzo}{2020}]{abdikamalov20acoustic}
Abdikamalov E.,  Foglizzo T.,  2020, \mn@doi [Royal Astronomical Society] {https://doi.org/10.1093/mnras/staa533}, 439, 3496

\bibitem[\protect\citeauthoryear{{Abdikamalov}, {Zhaksylykov}, {Radice}  \& {Berdibek}}{{Abdikamalov} et~al.}{2016}]{abdikamalov16shock}
{Abdikamalov} E.,  {Zhaksylykov} A.,  {Radice} D.,   {Berdibek} S.,  2016, \mn@doi [\mnras] {10.1093/mnras/stw1604}, \href {https://ui.adsabs.harvard.edu/abs/2016MNRAS.461.3864A} {461, 3864}

\bibitem[\protect\citeauthoryear{{Abdikamalov}, {Huete}, {Nussupbekov}  \& {Berdibek}}{{Abdikamalov} et~al.}{2018}]{abdikamalov18turbulence}
{Abdikamalov} E.,  {Huete} C.,  {Nussupbekov} A.,   {Berdibek} S.,  2018, \mn@doi [Particles] {10.3390/particles1010007}, \href {https://ui.adsabs.harvard.edu/abs/2018Parti...1....7A} {1, 7}

\bibitem[\protect\citeauthoryear{{Abdikamalov}, {Foglizzo}  \& {Mukazhanov}}{{Abdikamalov} et~al.}{2021}]{abdikamalov21impact}
{Abdikamalov} E.,  {Foglizzo} T.,   {Mukazhanov} O.,  2021, \mn@doi [\mnras] {10.1093/mnras/stab715}, \href {https://ui.adsabs.harvard.edu/abs/2021MNRAS.503.3617A} {503, 3617}

\bibitem[\protect\citeauthoryear{{Blondin}, {Mezzacappa}  \& {DeMarino}}{{Blondin} et~al.}{2003}]{blondin03stability}
{Blondin} J.~M.,  {Mezzacappa} A.,   {DeMarino} C.,  2003, \mn@doi [\apj] {10.1086/345812}, \href {https://ui.adsabs.harvard.edu/abs/2003ApJ...584..971B} {584, 971}

\bibitem[\protect\citeauthoryear{{Bondi}}{{Bondi}}{1952}]{bondi52spherically}
{Bondi} H.,  1952, \mn@doi [\mnras] {10.1093/mnras/112.2.195}, \href {https://ui.adsabs.harvard.edu/abs/1952MNRAS.112..195B} {112, 195}

\bibitem[\protect\citeauthoryear{{Bruenn} et~al.,}{{Bruenn} et~al.}{2016}]{bruenn16development}
{Bruenn} S.~W.,  et~al., 2016, \mn@doi [\apj] {10.3847/0004-637X/818/2/123}, \href {https://ui.adsabs.harvard.edu/abs/2016ApJ...818..123B} {818, 123}

\bibitem[\protect\citeauthoryear{{Burrows}}{{Burrows}}{2013}]{burrows13colloquium}
{Burrows} A.,  2013, \mn@doi [Reviews of Modern Physics] {10.1103/RevModPhys.85.245}, \href {https://ui.adsabs.harvard.edu/abs/2013RvMP...85..245B} {85, 245}

\bibitem[\protect\citeauthoryear{{Burrows}, {Hayes}  \& {Fryxell}}{{Burrows} et~al.}{1995}]{burrows95on}
{Burrows} A.,  {Hayes} J.,   {Fryxell} B.~A.,  1995, \mn@doi [\apj] {10.1086/176188}, \href {https://ui.adsabs.harvard.edu/abs/1995ApJ...450..830B} {450, 830}

\bibitem[\protect\citeauthoryear{{Burrows}, {Radice}  \& {Vartanyan}}{{Burrows} et~al.}{2019}]{burrows19three}
{Burrows} A.,  {Radice} D.,   {Vartanyan} D.,  2019, \mn@doi [\mnras] {10.1093/mnras/stz543}, \href {https://ui.adsabs.harvard.edu/abs/2019MNRAS.485.3153B} {485, 3153}

\bibitem[\protect\citeauthoryear{{\v{C}}ada \& Torrilhon}{{\v{C}}ada \& Torrilhon}{2009}]{vcada2009compact}
{\v{C}}ada M.,  Torrilhon M.,  2009, Journal of Computational Physics, 228, 4118

\bibitem[\protect\citeauthoryear{{Cantiello}, {Mankovich}, {Bildsten}, {Christensen-Dalsgaard}  \& {Paxton}}{{Cantiello} et~al.}{2014}]{cantiello14angular}
{Cantiello} M.,  {Mankovich} C.,  {Bildsten} L.,  {Christensen-Dalsgaard} J.,   {Paxton} B.,  2014, \mn@doi [\apj] {10.1088/0004-637X/788/1/93}, \href {https://ui.adsabs.harvard.edu/abs/2014ApJ...788...93C} {788, 93}

\bibitem[\protect\citeauthoryear{{Chatzopoulos}, {Graziani}  \& {Couch}}{{Chatzopoulos} et~al.}{2014}]{chatzopoulos14characterizing}
{Chatzopoulos} E.,  {Graziani} C.,   {Couch} S.~M.,  2014, \mn@doi [\apj] {10.1088/0004-637X/795/1/92}, \href {https://ui.adsabs.harvard.edu/abs/2014ApJ...795...92C} {795, 92}

\bibitem[\protect\citeauthoryear{{Collins}, {M{\"u}ller}  \& {Heger}}{{Collins} et~al.}{2018}]{collions18properties}
{Collins} C.,  {M{\"u}ller} B.,   {Heger} A.,  2018, \mn@doi [\mnras] {10.1093/mnras/stx2470}, \href {https://ui.adsabs.harvard.edu/abs/2018MNRAS.473.1695C} {473, 1695}

\bibitem[\protect\citeauthoryear{{Couch} \& {Ott}}{{Couch} \& {Ott}}{2013}]{couch13revival}
{Couch} S.~M.,  {Ott} C.~D.,  2013, \mn@doi [\apjl] {10.1088/2041-8205/778/1/L7}, \href {https://ui.adsabs.harvard.edu/abs/2013ApJ...778L...7C} {778, L7}

\bibitem[\protect\citeauthoryear{{Couch}, {Chatzopoulos}, {Arnett}  \& {Timmes}}{{Couch} et~al.}{2015}]{couch15three}
{Couch} S.~M.,  {Chatzopoulos} E.,  {Arnett} W.~D.,   {Timmes} F.~X.,  2015, \mn@doi [\apjl] {10.1088/2041-8205/808/1/L21}, \href {https://ui.adsabs.harvard.edu/abs/2015ApJ...808L..21C} {808, L21}

\bibitem[\protect\citeauthoryear{{Deheuvels} et~al.,}{{Deheuvels} et~al.}{2014}]{deheuvels14seismic}
{Deheuvels} S.,  et~al., 2014, \mn@doi [\aap] {10.1051/0004-6361/201322779}, \href {https://ui.adsabs.harvard.edu/abs/2014A&A...564A..27D} {564, A27}

\bibitem[\protect\citeauthoryear{{Endeve}, {Cardall}, {Budiardja}, {Beck}, {Bejnood}, {Toedte}, {Mezzacappa}  \& {Blondin}}{{Endeve} et~al.}{2012}]{endeve12turbulent}
{Endeve} E.,  {Cardall} C.~Y.,  {Budiardja} R.~D.,  {Beck} S.~W.,  {Bejnood} A.,  {Toedte} R.~J.,  {Mezzacappa} A.,   {Blondin} J.~M.,  2012, \mn@doi [\apj] {10.1088/0004-637X/751/1/26}, \href {https://ui.adsabs.harvard.edu/abs/2012ApJ...751...26E} {751, 26}

\bibitem[\protect\citeauthoryear{{Ertl}, {Janka}, {Woosley}, {Sukhbold}  \& {Ugliano}}{{Ertl} et~al.}{2016}]{ertl16two}
{Ertl} T.,  {Janka} H.~T.,  {Woosley} S.~E.,  {Sukhbold} T.,   {Ugliano} M.,  2016, \mn@doi [\apj] {10.3847/0004-637X/818/2/124}, \href {https://ui.adsabs.harvard.edu/abs/2016ApJ...818..124E} {818, 124}

\bibitem[\protect\citeauthoryear{{Fern{\'a}ndez}}{{Fern{\'a}ndez}}{2015}]{fernandez15three}
{Fern{\'a}ndez} R.,  2015, \mn@doi [\mnras] {10.1093/mnras/stv1463}, \href {https://ui.adsabs.harvard.edu/abs/2015MNRAS.452.2071F} {452, 2071}

\bibitem[\protect\citeauthoryear{{Fields} \& {Couch}}{{Fields} \& {Couch}}{2021}]{fields21three}
{Fields} C.~E.,  {Couch} S.~M.,  2021, \mn@doi [\apj] {10.3847/1538-4357/ac24fb}, \href {https://ui.adsabs.harvard.edu/abs/2021ApJ...921...28F} {921, 28}

\bibitem[\protect\citeauthoryear{{Foglizzo}}{{Foglizzo}}{2001}]{foglizzo01entropy}
{Foglizzo} T.,  2001, \mn@doi [\aap] {10.1051/0004-6361:20000506}, \href {https://ui.adsabs.harvard.edu/abs/2001A&A...368..311F} {368, 311}

\bibitem[\protect\citeauthoryear{{Foglizzo}}{{Foglizzo}}{2002}]{foglizzo02nonradial}
{Foglizzo} T.,  2002, \mn@doi [\aap] {10.1051/0004-6361:20020912}, \href {https://ui.adsabs.harvard.edu/abs/2002A&A...392..353F} {392, 353}

\bibitem[\protect\citeauthoryear{{Foglizzo}, {Galletti}, {Scheck}  \& {Janka}}{{Foglizzo} et~al.}{2007}]{foglizzo07sasi}
{Foglizzo} T.,  {Galletti} P.,  {Scheck} L.,   {Janka} H.~T.,  2007, \mn@doi [\apj] {10.1086/509612}, \href {https://ui.adsabs.harvard.edu/abs/2007ApJ...654.1006F} {654, 1006}

\bibitem[\protect\citeauthoryear{{Georgy} et~al.,}{{Georgy} et~al.}{2024}]{georgy243D}
{Georgy} C.,  et~al., 2024, \mn@doi [\mnras] {10.1093/mnras/stae1381}, \href {https://ui.adsabs.harvard.edu/abs/2024MNRAS.531.4293G} {531, 4293}

\bibitem[\protect\citeauthoryear{{Heger}, {Woosley}  \& {Spruit}}{{Heger} et~al.}{2005}]{heger05presupernova}
{Heger} A.,  {Woosley} S.~E.,   {Spruit} H.~C.,  2005, \mn@doi [\apj] {10.1086/429868}, \href {https://ui.adsabs.harvard.edu/abs/2005ApJ...626..350H} {626, 350}

\bibitem[\protect\citeauthoryear{{Herant}, {Benz}, {Hix}, {Fryer}  \& {Colgate}}{{Herant} et~al.}{1994}]{herant94inside}
{Herant} M.,  {Benz} W.,  {Hix} W.~R.,  {Fryer} C.~L.,   {Colgate} S.~A.,  1994, \mn@doi [\apj] {10.1086/174817}, \href {https://ui.adsabs.harvard.edu/abs/1994ApJ...435..339H} {435, 339}

\bibitem[\protect\citeauthoryear{{Huete}, {Abdikamalov}  \& {Radice}}{{Huete} et~al.}{2018}]{huete18impact}
{Huete} C.,  {Abdikamalov} E.,   {Radice} D.,  2018, \mn@doi [\mnras] {10.1093/mnras/stx3360}, \href {https://ui.adsabs.harvard.edu/abs/2018MNRAS.475.3305H} {475, 3305}

\bibitem[\protect\citeauthoryear{{Iwakami}, {Kotake}, {Ohnishi}, {Yamada}  \& {Sawada}}{{Iwakami} et~al.}{2008}]{iwakami08three}
{Iwakami} W.,  {Kotake} K.,  {Ohnishi} N.,  {Yamada} S.,   {Sawada} K.,  2008, \mn@doi [\apj] {10.1086/533582}, \href {https://ui.adsabs.harvard.edu/abs/2008ApJ...678.1207I} {678, 1207}

\bibitem[\protect\citeauthoryear{{Janka}}{{Janka}}{2001}]{janka01conditions}
{Janka} H.~T.,  2001, \mn@doi [\aap] {10.1051/0004-6361:20010012}, \href {https://ui.adsabs.harvard.edu/abs/2001A&A...368..527J} {368, 527}

\bibitem[\protect\citeauthoryear{{Janka} \& {Mueller}}{{Janka} \& {Mueller}}{1995}]{janka95first}
{Janka} H.~T.,  {Mueller} E.,  1995, \mn@doi [\apjl] {10.1086/309604}, \href {https://ui.adsabs.harvard.edu/abs/1995ApJ...448L.109J} {448, L109}

\bibitem[\protect\citeauthoryear{{Janka}, {Melson}  \& {Summa}}{{Janka} et~al.}{2016}]{janka16physics}
{Janka} H.-T.,  {Melson} T.,   {Summa} A.,  2016, \mn@doi [Annual Review of Nuclear and Particle Science] {10.1146/annurev-nucl-102115-044747}, \href {https://ui.adsabs.harvard.edu/abs/2016ARNPS..66..341J} {66, 341}

\bibitem[\protect\citeauthoryear{{Kazeroni} \& {Abdikamalov}}{{Kazeroni} \& {Abdikamalov}}{2020}]{kazeroni20impact}
{Kazeroni} R.,  {Abdikamalov} E.,  2020, \mn@doi [\mnras] {10.1093/mnras/staa944}, \href {https://ui.adsabs.harvard.edu/abs/2020MNRAS.494.5360K} {494, 5360}

\bibitem[\protect\citeauthoryear{{Kovalenko} \& {Eremin}}{{Kovalenko} \& {Eremin}}{1998}]{kovalenko98instability}
{Kovalenko} I.~G.,  {Eremin} M.~A.,  1998, \mn@doi [\mnras] {10.1046/j.1365-8711.1998.01667.x}, \href {https://ui.adsabs.harvard.edu/abs/1998MNRAS.298..861K} {298, 861}

\bibitem[\protect\citeauthoryear{{Lai} \& {Goldreich}}{{Lai} \& {Goldreich}}{2000}]{lai00growth}
{Lai} D.,  {Goldreich} P.,  2000, \mn@doi [\apj] {10.1086/308821}, \href {https://ui.adsabs.harvard.edu/abs/2000ApJ...535..402L} {535, 402}

\bibitem[\protect\citeauthoryear{{Landau} \& {Lifshitz}}{{Landau} \& {Lifshitz}}{1987}]{landau87fluid}
{Landau} L.~D.,  {Lifshitz} E.~M.,  1987, {Fluid Mechanics}

\bibitem[\protect\citeauthoryear{{Lesur}, {Baghdadi}, {Wafflard-Fernandez}, {Mauxion}, {Robert}  \& {Van den Bossche}}{{Lesur} et~al.}{2023}]{lesur23idefix}
{Lesur} G.~R.~J.,  {Baghdadi} S.,  {Wafflard-Fernandez} G.,  {Mauxion} J.,  {Robert} C.~M.~T.,   {Van den Bossche} M.,  2023, \mn@doi [\aap] {10.1051/0004-6361/202346005}, \href {https://ui.adsabs.harvard.edu/abs/2023A&A...677A...9L} {677, A9}

\bibitem[\protect\citeauthoryear{{Liebend{\"o}rfer}, {Mezzacappa}, {Thielemann}, {Messer}, {Hix}  \& {Bruenn}}{{Liebend{\"o}rfer} et~al.}{2001}]{liebendoerfer01probing}
{Liebend{\"o}rfer} M.,  {Mezzacappa} A.,  {Thielemann} F.-K.,  {Messer} O.~E.,  {Hix} W.~R.,   {Bruenn} S.~W.,  2001, \mn@doi [\prd] {10.1103/PhysRevD.63.103004}, \href {https://ui.adsabs.harvard.edu/abs/2001PhRvD..63j3004L} {63, 103004}

\bibitem[\protect\citeauthoryear{{Matsumoto}, {Asahina}, {Takiwaki}, {Kotake}  \& {Takahashi}}{{Matsumoto} et~al.}{2022}]{matsumoto22magnetic}
{Matsumoto} J.,  {Asahina} Y.,  {Takiwaki} T.,  {Kotake} K.,   {Takahashi} H.~R.,  2022, \mn@doi [\mnras] {10.1093/mnras/stac2335}, \href {https://ui.adsabs.harvard.edu/abs/2022MNRAS.516.1752M} {516, 1752}

\bibitem[\protect\citeauthoryear{{Melson}, {Kresse}  \& {Janka}}{{Melson} et~al.}{2020}]{melson20resolution}
{Melson} T.,  {Kresse} D.,   {Janka} H.-T.,  2020, \mn@doi [\apj] {10.3847/1538-4357/ab72a7}, \href {https://ui.adsabs.harvard.edu/abs/2020ApJ...891...27M} {891, 27}

\bibitem[\protect\citeauthoryear{{Mezzacappa}, {Endeve}, {Messer}  \& {Bruenn}}{{Mezzacappa} et~al.}{2020}]{mezzacappa20physical}
{Mezzacappa} A.,  {Endeve} E.,  {Messer} O.~E.~B.,   {Bruenn} S.~W.,  2020, \mn@doi [Living Reviews in Computational Astrophysics] {10.1007/s41115-020-00010-8}, \href {https://ui.adsabs.harvard.edu/abs/2020LRCA....6....4M} {6, 4}

\bibitem[\protect\citeauthoryear{{M{\"u}ller}}{{M{\"u}ller}}{2020}]{muller20hydrodynamics}
{M{\"u}ller} B.,  2020, \mn@doi [Living Reviews in Computational Astrophysics] {10.1007/s41115-020-0008-5}, \href {https://ui.adsabs.harvard.edu/abs/2020LRCA....6....3M} {6, 3}

\bibitem[\protect\citeauthoryear{M{\"u}ller \& Janka}{M{\"u}ller \& Janka}{2015}]{muller15non}
M{\"u}ller B.,  Janka H.-T.,  2015, \mn@doi [Royal Astronomical Society] {https://doi.org/10.1093/mnras/stv101}, 448, 2141

\bibitem[\protect\citeauthoryear{{M{\"u}ller}, {Viallet}, {Heger}  \& {Janka}}{{M{\"u}ller} et~al.}{2016}]{muller16last}
{M{\"u}ller} B.,  {Viallet} M.,  {Heger} A.,   {Janka} H.-T.,  2016, \mn@doi [\apj] {10.3847/1538-4357/833/1/124}, \href {https://ui.adsabs.harvard.edu/abs/2016ApJ...833..124M} {833, 124}

\bibitem[\protect\citeauthoryear{{M{\"u}ller}, {Melson}, {Heger}  \& {Janka}}{{M{\"u}ller} et~al.}{2017}]{muller17supernova}
{M{\"u}ller} B.,  {Melson} T.,  {Heger} A.,   {Janka} H.-T.,  2017, \mn@doi [\mnras] {10.1093/mnras/stx1962}, \href {https://ui.adsabs.harvard.edu/abs/2017MNRAS.472..491M} {472, 491}

\bibitem[\protect\citeauthoryear{{Murphy}, {Dolence}  \& {Burrows}}{{Murphy} et~al.}{2013}]{murphy13dominance}
{Murphy} J.~W.,  {Dolence} J.~C.,   {Burrows} A.,  2013, \mn@doi [\apj] {10.1088/0004-637X/771/1/52}, \href {https://ui.adsabs.harvard.edu/abs/2013ApJ...771...52M} {771, 52}

\bibitem[\protect\citeauthoryear{{Noutsos}, {Schnitzeler}, {Keane}, {Kramer}  \& {Johnston}}{{Noutsos} et~al.}{2013}]{noutsos13}
{Noutsos} A.,  {Schnitzeler} D.~H.~F.~M.,  {Keane} E.~F.,  {Kramer} M.,   {Johnston} S.,  2013, \mn@doi [\mnras] {10.1093/mnras/stt047}, \href {https://ui.adsabs.harvard.edu/abs/2013MNRAS.430.2281N} {430, 2281}

\bibitem[\protect\citeauthoryear{{O'Connor} \& {Ott}}{{O'Connor} \& {Ott}}{2011}]{oconnor11black}
{O'Connor} E.,  {Ott} C.~D.,  2011, \mn@doi [\apj] {10.1088/0004-637X/730/2/70}, \href {https://ui.adsabs.harvard.edu/abs/2011ApJ...730...70O} {730, 70}

\bibitem[\protect\citeauthoryear{{Popov} \& {Turolla}}{{Popov} \& {Turolla}}{2012}]{popov12}
{Popov} S.~B.,  {Turolla} R.,  2012, \mn@doi [\apss] {10.1007/s10509-012-1100-z}, \href {https://ui.adsabs.harvard.edu/abs/2012Ap&SS.341..457P} {341, 457}

\bibitem[\protect\citeauthoryear{{Radice}, {Ott}, {Abdikamalov}, {Couch}, {Haas}  \& {Schnetter}}{{Radice} et~al.}{2016}]{radice16neutrino}
{Radice} D.,  {Ott} C.~D.,  {Abdikamalov} E.,  {Couch} S.~M.,  {Haas} R.,   {Schnetter} E.,  2016, \mn@doi [\apj] {10.3847/0004-637X/820/1/76}, \href {https://ui.adsabs.harvard.edu/abs/2016ApJ...820...76R} {820, 76}

\bibitem[\protect\citeauthoryear{{Radice}, {Burrows}, {Vartanyan}, {Skinner}  \& {Dolence}}{{Radice} et~al.}{2017}]{radice17electron}
{Radice} D.,  {Burrows} A.,  {Vartanyan} D.,  {Skinner} M.~A.,   {Dolence} J.~C.,  2017, \mn@doi [\apj] {10.3847/1538-4357/aa92c5}, \href {https://ui.adsabs.harvard.edu/abs/2017ApJ...850...43R} {850, 43}

\bibitem[\protect\citeauthoryear{{Radice}, {Abdikamalov}, {Ott}, {M{\"o}sta}, {Couch}  \& {Roberts}}{{Radice} et~al.}{2018}]{radice18turbulence}
{Radice} D.,  {Abdikamalov} E.,  {Ott} C.~D.,  {M{\"o}sta} P.,  {Couch} S.~M.,   {Roberts} L.~F.,  2018, \mn@doi [Journal of Physics G Nuclear Physics] {10.1088/1361-6471/aab872}, \href {https://ui.adsabs.harvard.edu/abs/2018JPhG...45e3003R} {45, 053003}

\bibitem[\protect\citeauthoryear{{Suwa}, {Yamada}, {Takiwaki}  \& {Kotake}}{{Suwa} et~al.}{2016}]{suwa16Acriterion}
{Suwa} Y.,  {Yamada} S.,  {Takiwaki} T.,   {Kotake} K.,  2016, \mn@doi [\apj] {10.3847/0004-637X/816/1/43}, \href {https://ui.adsabs.harvard.edu/abs/2016ApJ...816...43S} {816, 43}

\bibitem[\protect\citeauthoryear{{Takahashi} \& {Yamada}}{{Takahashi} \& {Yamada}}{2014}]{takahashi14linear}
{Takahashi} K.,  {Yamada} S.,  2014, \mn@doi [\apj] {10.1088/0004-637X/794/2/162}, \href {https://ui.adsabs.harvard.edu/abs/2014ApJ...794..162T} {794, 162}

\bibitem[\protect\citeauthoryear{{Takahashi}, {Iwakami}, {Yamamoto}  \& {Yamada}}{{Takahashi} et~al.}{2016}]{takahashi16links}
{Takahashi} K.,  {Iwakami} W.,  {Yamamoto} Y.,   {Yamada} S.,  2016, \mn@doi [\apj] {10.3847/0004-637X/831/1/75}, \href {https://ui.adsabs.harvard.edu/abs/2016ApJ...831...75T} {831, 75}

\bibitem[\protect\citeauthoryear{{Ugliano}, {Janka}, {Marek}  \& {Arcones}}{{Ugliano} et~al.}{2012}]{ugliano12progenitor}
{Ugliano} M.,  {Janka} H.-T.,  {Marek} A.,   {Arcones} A.,  2012, \mn@doi [\apj] {10.1088/0004-637X/757/1/69}, \href {https://ui.adsabs.harvard.edu/abs/2012ApJ...757...69U} {757, 69}

\bibitem[\protect\citeauthoryear{{Varma} \& {M{\"u}ller}}{{Varma} \& {M{\"u}ller}}{2023}]{varma233D}
{Varma} V.,  {M{\"u}ller} B.,  2023, \mn@doi [\mnras] {10.1093/mnras/stad3113}, \href {https://ui.adsabs.harvard.edu/abs/2023MNRAS.526.5249V} {526, 5249}

\bibitem[\protect\citeauthoryear{{Yadav}, {M{\"u}ller}, {Janka}, {Melson}  \& {Heger}}{{Yadav} et~al.}{2020}]{yadav20large}
{Yadav} N.,  {M{\"u}ller} B.,  {Janka} H.~T.,  {Melson} T.,   {Heger} A.,  2020, \mn@doi [\apj] {10.3847/1538-4357/ab66bb}, \href {https://ui.adsabs.harvard.edu/abs/2020ApJ...890...94Y} {890, 94}

\bibitem[\protect\citeauthoryear{{Yoshida}, {Takiwaki}, {Aguilera-Dena}, {Kotake}, {Takahashi}, {Nakamura}, {Umeda}  \& {Langer}}{{Yoshida} et~al.}{2021}]{yoshida21three}
{Yoshida} T.,  {Takiwaki} T.,  {Aguilera-Dena} D.~R.,  {Kotake} K.,  {Takahashi} K.,  {Nakamura} K.,  {Umeda} H.,   {Langer} N.,  2021, \mn@doi [\mnras] {10.1093/mnrasl/slab067}, \href {https://ui.adsabs.harvard.edu/abs/2021MNRAS.506L..20Y} {506, L20}

\makeatother
\end{thebibliography}

\bsp	
\label{lastpage}
\end{document}